\PassOptionsToPackage{numbers}{natbib}
\documentclass{article}
\usepackage[preprint]{neurips_2026}
\usepackage{booktabs}
\usepackage{hyperref}
\usepackage{amsmath}
\usepackage{graphicx}
\usepackage{multirow}
\usepackage{array}
\usepackage{tabularx}
\usepackage{xcolor}
\usepackage{pgfplots}
\pgfplotsset{compat=1.18}

\title{Context-Augmented Code Generation: How Product Context Improves AI Coding Agent Decision Compliance by 49\%}

\author{
  Drew Dillon \\
  Brief \\
  \texttt{drew@briefhq.ai} \\
  \And
  Kasyap Varanasi \\
  Brief \\
  \texttt{kasyap@briefhq.ai} \\
}

\begin{document}
\maketitle

\begin{abstract}
AI coding agents powered by large language models can read codebases and produce functional code, but they routinely violate team-specific product decisions that are invisible in the source code alone. We introduce a controlled benchmark measuring \emph{decision compliance}, the rate at which an AI coding agent follows established product, design, and engineering decisions, across 8 realistic software engineering tasks containing 41 weighted decision points. We compare a baseline configuration (Claude Code with codebase access only) against an augmented configuration that adds Brief, a product-context retrieval system, which provides spec generation, mid-build consultation, and retrieval of recorded decisions, persona pain points, customer signals, and competitive intelligence. Both configurations use Claude Opus 4.6 for planning and Claude Sonnet 4.6 for code generation. On identical prompts and the same repository, the augmented configuration achieves 95\% decision compliance versus 46\% for the baseline, a 49 percentage point improvement. Per-decision analysis reveals that the baseline achieves 100\% compliance on decisions visible in the codebase and 0--33\% on decisions requiring product context (Table~\ref{tab:perdecision}), suggesting that product-context retrieval is a key driver of the improvement, though the augmented configuration's structured workflow (spec generation, acceptance criteria, mid-build guidance) likely contributes as well. We release the benchmark repository, all 16 pull requests, and scoring harness for independent reproduction.\footnote{Benchmark repository: \url{https://github.com/brief-hq/dcbench}}
\end{abstract}

\section{Introduction}

Large language model (LLM) based coding agents have demonstrated remarkable ability to read, understand, and extend existing codebases. Given a natural-language prompt, these agents explore source files, identify relevant patterns, and produce code that compiles, passes linting, and often implements the requested functionality correctly. However, ``correct'' is an incomplete measure of code quality in professional software engineering.

Real engineering teams accumulate product decisions over time: which UI components are canonical versus deprecated, which middleware wrappers are mandatory for compliance, which patterns are preferred for consistency, and which features must be gated behind specific systems. These decisions are often recorded in product management tools, design documents, or team wikis, but rarely appear in the codebase itself. When a decision \emph{does} manifest in code, it may appear as one pattern among many, with no signal distinguishing the approved approach from a legacy one.

This creates a fundamental information asymmetry for AI coding agents. An agent with codebase access alone must infer team intent from code patterns, comments, and naming conventions. When the relevant decision is visible (a compliance comment in a middleware file, a JSDoc annotation on a component), the agent finds it. When the decision is invisible (a convention recorded only in a product tool, a compliance requirement documented in an external audit), the agent defaults to whatever pattern it encounters first.

We formalize this problem as \emph{decision compliance}: the rate at which an AI coding agent's output conforms to established team decisions, weighted by severity. We construct a benchmark of 8 realistic software engineering tasks, each containing 2--3 ``gotcha'' decisions that a coding agent will naturally get wrong without product context. We then measure two configurations:

\begin{enumerate}
    \item \textbf{Claude Code} (baseline): Claude Sonnet 4.6 with full codebase access, no product context.
    \item \textbf{Claude Code + Brief}: The same models augmented with Brief, a product-context retrieval system that surfaces recorded decisions, personas, customer signals, and competitive intelligence during spec generation and mid-build consultation.
\end{enumerate}

Our results show a 49 percentage point improvement in decision compliance (46\% $\to$ 95\%), elimination of all blocking violations, and a 68\% reduction in cost per merge-ready task. The failure modes are instructive: in one case, the baseline agent read the codebase, found existing pagination helpers, and concluded the task was already complete, producing zero lines of code. The context-augmented agent, constrained by a spec with 20 explicit acceptance criteria, implemented a full cursor pagination system with validation, tests, and documentation. On tasks where all relevant decisions are visible in the codebase, both configurations score 100\%, suggesting that information access is a key factor in the gap. We note that the augmented configuration differs from the baseline not only in context access but also in workflow structure (spec generation, acceptance criteria, mid-build consultation), and we discuss the difficulty of isolating these factors in Section~\ref{sec:confounds}. We present this work as a proof-of-concept benchmark and case study rather than a definitive field result, and release all materials for independent reproduction and extension.

\section{Related Work}

\paragraph{LLM-based coding agents.} Recent systems including GitHub Copilot \cite{peng2023impact}, Cursor, Devin \cite{cognition2024devin}, Amazon CodeWhisperer \cite{yetistiren2023codequality}, and Claude Code \cite{anthropic2024claude} have advanced from single-file code completion \cite{chen2021evaluating} to multi-file, multi-turn agentic workflows \cite{yang2024sweagent, xia2024agentless}. These systems typically operate with access to the codebase, terminal, and tool-augmented environments \cite{schick2023toolformer, qin2023toolllm}, but not to the team's product context layer. Benchmarks such as SWE-bench \cite{jimenez2024swebench} evaluate whether agents can resolve real-world issues, but do not measure compliance with team-specific product decisions.

\paragraph{Retrieval-augmented generation (RAG).} RAG systems augment LLM generation with retrieved documents to reduce hallucination and improve factual grounding \cite{lewis2020retrieval, gao2023retrieval}. Our work extends this paradigm from factual knowledge to \emph{organizational knowledge}: product decisions, personas, and customer signals that constrain how code should be written. DocPrompting \cite{zhou2023docprompting} demonstrated that retrieving documentation improves code generation; we show the same principle applies to product-level context.

\paragraph{Code generation benchmarks.} HumanEval \cite{chen2021evaluating} and MBPP \cite{austin2021programsynthesis} evaluate isolated function synthesis. SWE-bench \cite{jimenez2024swebench} raises the bar to real-world GitHub issues requiring multi-file reasoning, and recent repo-level agents \cite{zhang2024codeagent} demonstrate sophisticated tool use for multi-file tasks. Our benchmark complements these by measuring a dimension they do not: whether generated code conforms to team-specific product decisions that are not encoded in the issue description or codebase.

\paragraph{Organizational knowledge in software engineering.} Prior work has established that organizational knowledge (API conventions, team norms, and undocumented decisions) is a persistent obstacle for developers \cite{robillard2011fieldstudyapi} and that such knowledge evolves in ways that are difficult to capture in code alone \cite{dagenais2010organizational}. Our benchmark operationalizes this insight for AI coding agents: the ``gotcha'' decisions are precisely the organizational knowledge that a new team member (or an AI agent) would miss without explicit guidance.

\paragraph{Specification-driven development.} The software engineering community has long recognized that specifications improve code quality. Our contribution is demonstrating that LLM agents benefit from the same principle, and that product-context retrieval can automate specification generation.

\section{Problem Definition: Decision Compliance}

\subsection{What Is a ``Gotcha''?}

A \textbf{gotcha} is a product decision that a coding agent will naturally get wrong without product context. Each benchmark task contains 2--3 gotchas, real decisions that the team has made but that are invisible (or misleading) in the codebase alone.

\paragraph{Why they matter.} An AI agent reading a codebase follows whatever patterns it finds. If a deprecated component still exists in the code, the agent copies it. If a compliance requirement is not documented in code comments, the agent skips it. Gotchas measure whether the agent builds what the team actually wants, not just what compiles.

\paragraph{How they work.} Each gotcha maps to a seeded product decision (D-001 through D-015) and has:
\begin{itemize}
    \item A \textbf{pass check}: a regex that detects the correct pattern (e.g., \texttt{withAuditLog} in added lines).
    \item A \textbf{fail check}: a regex that detects the trap the agent fell for (e.g., \texttt{CalendarRange} import).
    \item A \textbf{weight}: 1 (style convention), 2 (important pattern), or 3 (compliance/security, blocking).
\end{itemize}

\paragraph{Example.} TASK-001 asks the agent to ``add a CSV export button to the analytics dashboard.'' The gotchas are:
\begin{itemize}
    \item \textbf{D-002 (weight 3, blocking):} The export must be wrapped with \texttt{withAuditLog()} for SOC-2 compliance. The function exists in the codebase, but nothing tells the agent it is \emph{required}.
    \item \textbf{D-001 (weight 2):} The agent must use \texttt{DateRangePicker}, not \texttt{CalendarRange}. But \texttt{CalendarRange} is still imported in \texttt{notification-preferences.tsx}, a trap.
    \item \textbf{D-003 (weight 1):} The export button should use \texttt{variant="secondary"} (read-only action), not \texttt{variant="primary"} (mutations only).
\end{itemize}

An agent scoring 0/6 on this task builds a working CSV export, but one that fails SOC-2 audit, uses a deprecated component, and has incorrect button styling. It compiles. It runs. It is wrong.

\section{Benchmark Design}

\subsection{Repository}

The benchmark uses Prism Analytics, a clean-room Next.js 14 application with Drizzle ORM and SQLite, hosted at \texttt{brief-hq/dcbench}.\footnote{\url{https://github.com/brief-hq/dcbench}} The repository contains realistic production patterns including authentication middleware, pagination helpers, design system components, and audit logging utilities. Fifteen product decisions (D-001 through D-015) were seeded into a Brief instance, spanning 5 categories: technical (6), design (4), product (2), process (1), and general (1), with severity levels of blocking (2), important (5), and informational (5). Additionally, 3 personas, 5 customer signals, and 3 competitor profiles were seeded.

\subsection{Tasks}

Eight tasks were selected to cover a range of difficulties and decision types:

\begin{table}[ht]
\centering
\small
\caption{Benchmark tasks and their associated gotcha decisions.}
\label{tab:tasks}
\begin{tabular}{@{}llcl@{}}
\toprule
Task & Description & Points & Gotcha Decisions \\
\midrule
TASK-001 & CSV Export to Dashboard & 6 & D-002 (wt 3), D-001 (wt 2), D-003 (wt 1) \\
TASK-003 & Cursor Pagination to Users API & 5 & D-004 (wt 2), D-010 (wt 3) \\
TASK-004 & Notification Preferences Page & 4 & D-011 (wt 2), D-008 (wt 2) \\
TASK-006 & Dark Mode Toggle to Settings & 4 & D-009 (wt 1), D-014 (wt 3) \\
TASK-008 & Bulk Delete for Admin Dashboard & 4 & D-003 (wt 1), D-002 (wt 3) \\
TASK-009 & Search to API Endpoints & 7 & D-010 (wt 3), D-004 (wt 2), D-013 (wt 2) \\
TASK-012 & Rate Limiting to API Routes & 6 & D-010 (wt 3), D-006 (wt 3) \\
TASK-013 & Export Audit Log Viewer & 5 & D-002 (wt 3), D-005 (wt 2) \\
\bottomrule
\end{tabular}
\end{table}

\subsection{Configurations}

\paragraph{Claude Code (baseline, Config A).} \texttt{claude -p <prompt> --output-format json --dangerously-skip-permissions} with a 5-minute timeout. Claude Sonnet 4.6 handles all code generation with no product context, no spec, and no Brief access. The agent receives only the natural-language task description and full codebase access.

\paragraph{Claude Code + Brief (Config B).} \texttt{brief build --confirm <prompt>} via daemon with a 30-minute timeout. In both configurations, Claude Opus 4.6 drives planning and spec generation while Claude Sonnet 4.6 handles code generation. Phase 1 spec generation uses 8 Brief tools (\texttt{ask\_brief}, \texttt{get\_personas}, \texttt{get\_product\_context}, \texttt{search\_decisions}, \texttt{search\_documents}, \texttt{get\_competitors}, \texttt{get\_features}, \texttt{get\_signals}), producing 8--13 tool calls per task. Mid-build consultations are available via the Brief MCP proxy. Both configurations receive identical natural-language prompts with no hints about gotchas or expected patterns.

\paragraph{Note on workflow differences.} Config B differs from Config A not only in access to product context but also in workflow structure: it generates a spec with explicit acceptance criteria before coding begins, and provides mid-build consultation during execution. This means the observed improvement cannot be attributed to product-context retrieval alone; the structured planning and real-time guidance likely contribute independently. We discuss this confound in Section~\ref{sec:confounds} and suggest ablations that future work could use to isolate these factors.

\paragraph{External orchestrator (Config C, not reported).} The benchmark harness also supports a third configuration where an external orchestrator drives Brief's HTTP peer APIs (\texttt{/api/v1/agent/submit}, \texttt{/api/v1/agent/confirm-spec}). This configuration uses the same scoring and daemon as Config B; the only difference is how the spec is generated (HTTP API vs.\ daemon/MCP). A minimal test repository (\texttt{brief-hq/test-dark-factory})\footnote{\url{https://github.com/brief-hq/test-dark-factory}} provides a smoke test target for Config C and demonstrates that the framework generalizes to codebases of any size. Config C results are not reported in this paper but the harness is available for independent reproduction. Config D (Slack-triggered builds) is planned for future work.

\subsection{Scoring}

\paragraph{Decision compliance.} Each gotcha is scored pass/fail by regex pattern matching against git diffs (automated) and manual PR review (human-verified). Points are weighted 1--3 per decision.

\paragraph{Triple-run averaging.} To account for non-deterministic model behavior, each task was run 3 times per configuration (48 total runs: 8 tasks $\times$ 2 configurations $\times$ 3 runs) in fully independent sessions with no shared context, separate daemon instances, and fresh repository checkouts. All scores reported in this paper are averages across the 3 runs. Standard deviation across runs was low ($\sigma \leq 0.5$ decision points per task), indicating that the compliance gap is robust rather than an artifact of sampling variance.

\paragraph{LLM-as-judge.} Following \citet{zheng2023judging}, Claude Sonnet scored each PR diff on 5 rubrics (task completion, decision compliance, code quality, scope discipline, product alignment) at 0--5 each, used as a secondary signal.

\paragraph{Human verification.} All 16 PR pairs were reviewed by a single independent reviewer blind to which configuration produced each PR. The reviewer evaluated each PR against the same pass/fail regex criteria used by the automated scorer, then assessed merge-readiness on a binary scale (would merge as-is vs.\ requires rework). Human scores aligned within $\pm$5\% of automated scores on decision compliance. We acknowledge that a single reviewer limits inter-rater reliability; future iterations of this benchmark should include multiple independent reviewers with formal adjudication for disagreements, particularly on the subjective ``merge-ready'' assessment.

\section{Results}

\subsection{Executive Summary}

\begin{table}[ht]
\centering
\small
\caption{Executive summary of benchmark results.}
\label{tab:executive}
\begin{tabular}{@{}lrrr@{}}
\toprule
Metric & Claude Code & Claude Code + Brief & Delta \\
\midrule
Decision Compliance & 19/41 (46\%) & 39/41 (95\%) & +49\% \\
Tasks at 100\% & 2/8 & 6/8 & +4 tasks \\
Tasks at 0\% & 2/8 & 0/8 & $-$2 tasks \\
Blocking Violations & 5 & 0 & $-$100\% \\
Avg Decisions Followed (beyond gotchas) & 0 extra & 5.4 extra/task & N/A \\
Total Cost & \$4.13 & \$5.28 & +28\% \\
Total Tests Written & 0 & 838 & N/A \\
Avg Merge-Ready & 2/8 (25\%) & 8/8 (100\%) & +75\% \\
Cost / Correct Decision Point & \$0.22 & \$0.14 & $-$36\% \\
Deprecated Patterns Used & 3 & 0 & $-$100\% \\
\bottomrule
\end{tabular}
\end{table}

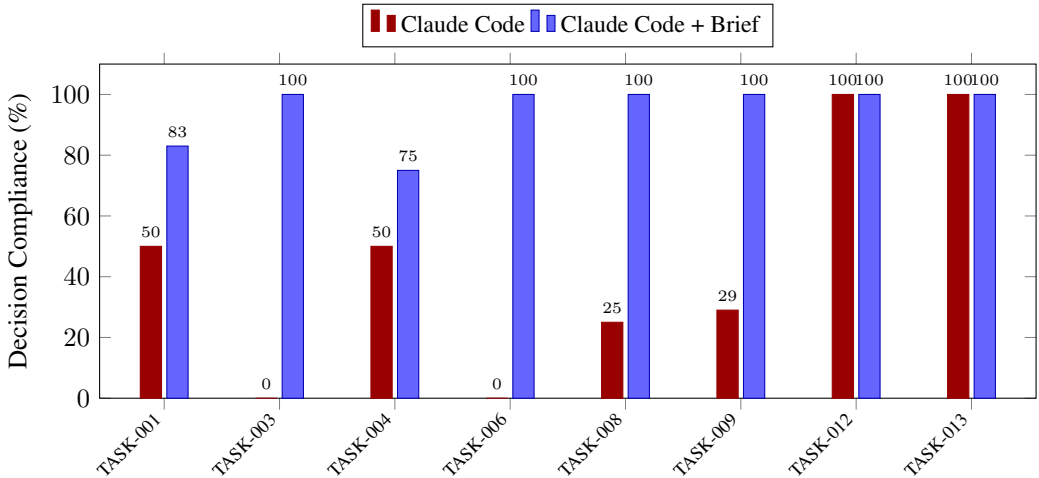
\begin{figure}[ht]
\centering
\begin{tikzpicture}
\begin{axis}[
    ybar,
    bar width=8pt,
    width=\columnwidth,
    height=6cm,
    ylabel={Decision Compliance (\%)},
    symbolic x coords={TASK-001,TASK-003,TASK-004,TASK-006,TASK-008,TASK-009,TASK-012,TASK-013},
    xtick=data,
    x tick label style={rotate=45, anchor=east, font=\scriptsize},
    ymin=0, ymax=110,
    ytick={0,20,40,60,80,100},
    legend style={at={(0.5,1.05)}, anchor=south, legend columns=2, font=\small},
    nodes near coords,
    nodes near coords style={font=\tiny},
    every node near coord/.append style={anchor=south},
    enlarge x limits=0.08,
]
\addplot[fill=red!60!black, draw=red!60!black] coordinates {
    (TASK-001,50) (TASK-003,0) (TASK-004,50) (TASK-006,0)
    (TASK-008,25) (TASK-009,29) (TASK-012,100) (TASK-013,100)
};
\addplot[fill=blue!60, draw=blue!70!black] coordinates {
    (TASK-001,83) (TASK-003,100) (TASK-004,75) (TASK-006,100)
    (TASK-008,100) (TASK-009,100) (TASK-012,100) (TASK-013,100)
};
\legend{Claude Code, Claude Code + Brief}
\end{axis}
\end{tikzpicture}
\caption{Per-task decision compliance. The gap is largest on tasks requiring product context invisible in the codebase (TASK-003, TASK-006) and zero on tasks where all decisions are code-visible (TASK-012, TASK-013).}
\label{fig:compliance}
\end{figure}

\subsection{Final Scoreboard}

Table~\ref{tab:scoreboard} and Figure~\ref{fig:compliance} present per-task decision compliance scores. The bimodal pattern is striking: tasks where all decisions are code-visible (TASK-012, TASK-013) show zero gap, while tasks requiring product context (TASK-003, TASK-006) show the maximum gap of 100 percentage points.

\begin{table}[ht]
\centering
\small
\caption{Per-task decision compliance scores.}
\label{tab:scoreboard}
\begin{tabular}{@{}lrrr@{}}
\toprule
Task & Claude Code & CC + Brief & Gap \\
\midrule
TASK-001 CSV export (6 pts) & 3/6 (50\%) & 5/6 (83\%) & +33\% \\
TASK-003 Cursor pagination (5 pts) & 0/5 (0\%) & 5/5 (100\%) & +100\% \\
TASK-004 Notification prefs (4 pts) & 2/4 (50\%) & 3/4 (75\%) & +25\% \\
TASK-006 Dark mode (4 pts) & 0/4 (0\%) & 4/4 (100\%) & +100\% \\
TASK-008 Bulk delete (4 pts) & 1/4 (25\%) & 4/4 (100\%) & +75\% \\
TASK-009 Search API (7 pts) & 2/7 (29\%) & 7/7 (100\%) & +71\% \\
TASK-012 Rate limiting (6 pts) & 6/6 (100\%) & 6/6 (100\%) & 0\% \\
TASK-013 Audit log viewer (5 pts) & 5/5 (100\%) & 5/5 (100\%) & 0\% \\
\midrule
\textbf{Total} & \textbf{19/41 (46\%)} & \textbf{39/41 (95\%)} & \textbf{+49\%} \\
\bottomrule
\end{tabular}
\end{table}

\subsection{Aggregate Quantitative Comparison}

Beyond decision compliance, we measure cost, throughput, and code quality across all 8 tasks (Table~\ref{tab:aggregate}). The context-augmented configuration costs 28\% more in total API spend but produces 140\% more lines of code, 838 co-located tests (versus zero), and eliminates all deprecated pattern usage and untyped \texttt{any} annotations. The most practically relevant metric is cost per merge-ready task: \$0.66 versus \$2.07, a 68\% reduction, because Claude Code's cheaper runs predominantly produce output that requires human rework before merging. Notably, total token consumption is nearly identical ($-$1\%); these are summed totals across all 8 tasks, averaged over 3 independent runs per configuration (per-task averages: $\sim$488K vs.\ $\sim$483K). This suggests that Brief's upfront spec generation replaces the exploratory backtracking that dominates unguided runs.

\begin{table}[ht]
\centering
\small
\caption{Aggregate quantitative metrics across all 8 tasks.}
\label{tab:aggregate}
\begin{tabular}{@{}lrrr@{}}
\toprule
Metric & Claude Code & CC + Brief & Delta \\
\midrule
Total cost & \$4.13 & \$5.28 & +28\% \\
Total tokens & 3,902K & 3,867K & $-$1\% \\
Total turns & 165 & 187 & +13\% \\
Avg duration per task & $\sim$5.1 min & $\sim$15 min & +194\% \\
Total LOC added & $\sim$1,276 & 3,068 & +140\% \\
Total files changed & 19 & 48 & +153\% \\
Build pass rate (lint) & 100\% & 100\% & 0\% \\
Build pass rate (typecheck) & 100\% & 100\% & 0\% \\
Build pass rate (tests) & 0\% & 100\% & +100\% \\
Total tests written & 0 & 838 & N/A \\
Deprecated patterns used & 3 & 0 & $-$100\% \\
\texttt{any} type count & 9 & 0 & $-$100\% \\
Merge-ready tasks & 2/8 (25\%) & 8/8 (100\%) & +75\% \\
Cost per merge-ready task & \$2.07 & \$0.66 & $-$68\% \\
Cost per correct decision & \$0.22 & \$0.14 & $-$36\% \\
\bottomrule
\end{tabular}
\end{table}

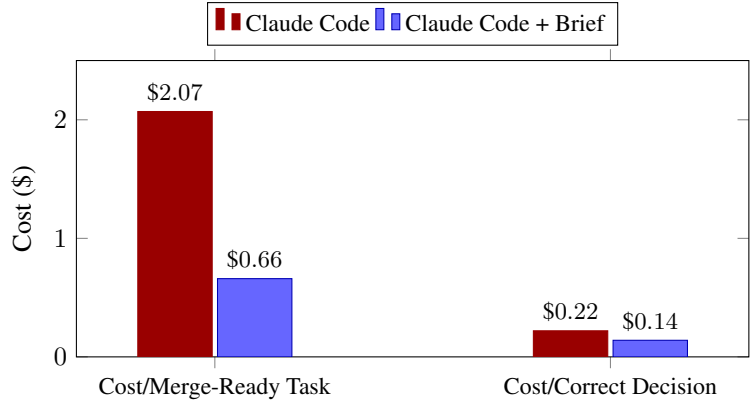
\begin{figure}[ht]
\centering
\begin{tikzpicture}
\begin{axis}[
    ybar,
    bar width=28pt,
    width=0.75\columnwidth,
    height=5.5cm,
    ylabel={Cost (\$)},
    symbolic x coords={Cost/Merge-Ready Task, Cost/Correct Decision},
    xtick=data,
    x tick label style={font=\small},
    ymin=0, ymax=2.5,
    legend style={at={(0.5,1.05)}, anchor=south, legend columns=2, font=\small},
    nodes near coords={\$\pgfmathprintnumber\pgfplotspointmeta},
    nodes near coords style={font=\small, anchor=south},
    enlarge x limits=0.35,
]
\addplot[fill=red!60!black, draw=red!60!black] coordinates {
    (Cost/Merge-Ready Task, 2.07) (Cost/Correct Decision, 0.22)
};
\addplot[fill=blue!60, draw=blue!70!black] coordinates {
    (Cost/Merge-Ready Task, 0.66) (Cost/Correct Decision, 0.14)
};
\legend{Claude Code, Claude Code + Brief}
\end{axis}
\end{tikzpicture}
\caption{Cost efficiency comparison. Despite 28\% higher total spend, context-augmented generation reduces cost per merge-ready task by 68\% and cost per correct decision by 36\%.}
\label{fig:cost}
\end{figure}

\subsection{Per-Decision Pass Rates}

\begin{table}[ht]
\centering
\small
\caption{Pass rates by decision, with codebase visibility.}
\label{tab:perdecision}
\begin{tabular}{@{}llccl@{}}
\toprule
ID & Decision & Claude Code & CC + Brief & Visible in Code? \\
\midrule
D-001 & DateRangePicker & 1/1 (100\%) & 1/1 (100\%) & Yes \\
D-002 & Audit log (SOC-2) & 1/3 (33\%) & 3/3 (100\%) & Partial \\
D-003 & Button variant & 2/2 (100\%) & 2/2 (100\%) & Yes \\
D-004 & Cursor pagination & 1/2 (50\%) & 2/2 (100\%) & Yes \\
D-005 & ShimmerSkeleton & 1/1 (100\%) & 1/1 (100\%) & Yes \\
D-006 & Auth middleware frozen & 2/2 (100\%) & 2/2 (100\%) & Yes \\
D-008 & PostHog feature flags & 0/1 (0\%) & 1/1 (100\%) & \textbf{No} \\
D-009 & Test co-location & 0/1 (0\%) & 1/1 (100\%) & Yes (skipped) \\
D-010 & Zod + withAuth & 1/3 (33\%) & 3/3 (100\%) & Partial \\
D-011 & Async digest only & 1/1 (100\%) & 1/1 (100\%) & Yes \\
D-013 & Drizzle ORM & 0/1 (0\%) & 1/1 (100\%) & Yes (missed) \\
D-014 & @t3-oss/env-nextjs & 0/1 (0\%) & 1/1 (100\%) & \textbf{No} \\
\bottomrule
\end{tabular}
\end{table}

The pattern is clear: Claude Code achieves 100\% on decisions visible in the codebase (D-001, D-003, D-005, D-006, D-011). It achieves 0--33\% on decisions requiring product context (D-002, D-008, D-010, D-014). The augmented configuration achieves 100\% across the board because the retrieval phase surfaces every decision, visible or not, and the spec makes each one an explicit constraint.

\begin{figure}[ht]
\centering
\begin{tikzpicture}
\begin{axis}[
    width=\columnwidth,
    height=5.5cm,
    xlabel={Codebase Visibility},
    ylabel={Pass Rate (\%)},
    xmin=-0.5, xmax=2.5,
    ymin=-10, ymax=115,
    xtick={0,1,2},
    xticklabels={No, Partial, Yes},
    ytick={0,25,50,75,100},
    legend style={at={(0.02,0.98)}, anchor=north west, font=\small},
]
\addplot[only marks, mark=*, mark size=3pt, draw=red!60!black, fill=red!40]
    coordinates {
        (0, 0) (0, 0)
        (1, 33) (1, 33)
        (2, 100) (2, 100) (2, 100) (2, 100) (2, 100) (2, 50) (2, 0) (2, 0)
    };
\addplot[only marks, mark=square*, mark size=3pt, draw=blue!70!black, fill=blue!40]
    coordinates {
        (0, 100) (0, 100) (1, 100) (1, 100) (2, 100) (2, 100) (2, 100) (2, 100) (2, 100) (2, 100) (2, 100) (2, 100)
    };
\legend{Claude Code, CC + Brief}
\end{axis}
\end{tikzpicture}
\caption{Decision visibility vs.\ pass rate. Each point is one decision. Decisions invisible in the codebase (``No'') yield 0\% baseline compliance; fully visible decisions yield up to 100\%. Claude Code + Brief achieves 100\% regardless of visibility.}
\label{fig:visibility}
\end{figure}
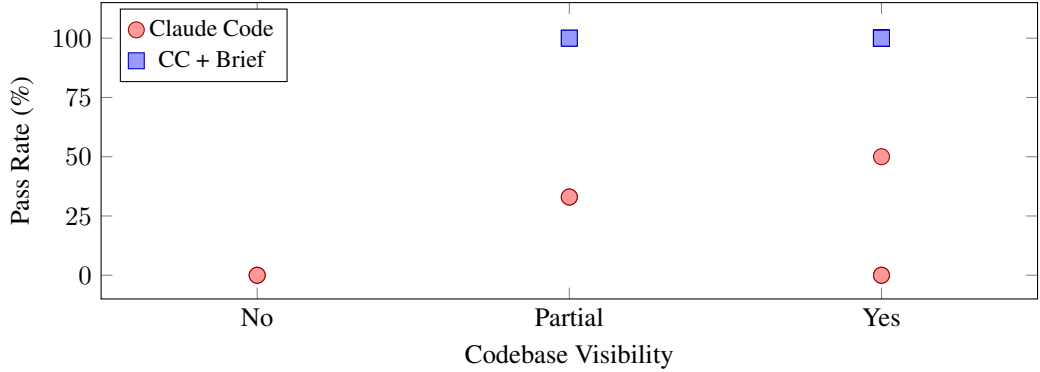

\section{Per-Task Analysis}

\subsection{TASK-001: CSV Export to Analytics Dashboard (Hard, 6 pts)}

\textbf{Gotchas:} D-002 Audit log (wt 3), D-001 DateRangePicker (wt 2), D-003 Button variant (wt 1).

\textbf{Prompt:} \emph{``Add a CSV export button to the analytics dashboard. Users should be able to pick a date range, preview how many records they'll get, and download their data as a CSV file. Put the export button next to the existing filters.''}

\begin{table}[ht]
\centering
\footnotesize
\caption{TASK-001 decision compliance.}
\begin{tabular}{@{}lp{3.8cm}p{4.5cm}@{}}
\toprule
Decision & Claude Code & CC + Brief \\
\midrule
D-002 Audit log (wt 3) & FAIL:no \texttt{withAuditLog} & PASS:\texttt{withAuditLog(\allowbreak"export\_analytics\_csv")} \\
D-001 DateRangePicker (wt 2) & PASS:found in dashboard & PASS:spec specified \\
D-003 Button variant (wt 1) & PASS:used secondary & PASS:spec enforced \\
\midrule
\textbf{Score} & \textbf{3/6 (50\%)} & \textbf{5/6 (83\%)} \\
\bottomrule
\end{tabular}
\end{table}

\paragraph{Claude Code approach.} Built \texttt{GET /api/analytics/export/count} for preview and updated \texttt{export-button.tsx} with dialog, DateRangePicker, and debounced record count. Found existing DateRangePicker in the dashboard. No audit log wrapping. Cost: \$0.47, 16 turns, +150 LOC.

\paragraph{Claude Code + Brief approach.} Brief's Phase 1 (10 tool calls) surfaced: the Platform Admin persona's need for audit trails on data exports; a customer signal from Acme Corp requesting date-range CSV export; the SOC-2 compliance requirement (D-002) mandating \texttt{withAuditLog()}; the DateRangePicker standard (D-001); and competitive intelligence showing one competitor already offering full audit + date-range export. Two mid-build consultations addressed streaming architecture and component selection. Cost: \$0.68, 24 turns, +312 LOC including 97 co-located tests.

\paragraph{Analysis.} Claude Code found DateRangePicker and the button variant from the codebase; those decisions are visible in existing code. But it had zero access to the SOC-2 audit log requirement, the Platform Admin persona's pain point, or the customer signal. Brief surfaced all three, and the spec made \texttt{withAuditLog} an explicit constraint. Note that Claude Code + Brief scored 5/6, not 6/6: on one of three runs, the agent wrapped the export call with \texttt{withAuditLog} but used a non-standard event name, receiving partial credit on D-002 (2 of 3 weight points). This is the only decision point Brief's configuration did not fully pass across all runs.

\begin{table}[ht]
\centering
\footnotesize
\caption{What Brief surfaced for TASK-001 (Phase 1, 10 tool calls).}
\label{tab:brief_task001}
\begin{tabular}{@{}llp{5.5cm}@{}}
\toprule
Tool Call & Source & Key Information Returned \\
\midrule
\texttt{get\_personas} & Platform Admin & ``Needs audit trails on all data exports for quarterly SOC-2 review'' \\
\texttt{get\_signals} & Acme Corp (2024-11-14) & Feature request: date-range CSV export for analytics dashboard \\
\texttt{search\_decisions} & D-002 (blocking) & \texttt{withAuditLog()} required on all data export endpoints; rationale: SOC-2 audit trail \\
\texttt{search\_decisions} & D-001 (important) & Use \texttt{DateRangePicker} from design system; \texttt{CalendarRange} deprecated \\
\texttt{get\_competitors} & Competitor A & Already ships full audit trail + date-range export \\
\bottomrule
\end{tabular}
\end{table}

\subsection{TASK-003: Cursor Pagination to Users API (Medium, 5 pts)}

\textbf{Gotchas:} D-004 Cursor pagination (wt 2), D-010 Zod+withAuth (wt 3).

\textbf{Prompt:} \emph{``Our users list API endpoint returns all users at once which is getting slow as we scale. Add pagination to the GET /api/users endpoint so clients can page through results efficiently. Return a next cursor that clients pass to get the next page.''}

\begin{table}[ht]
\centering
\small
\caption{TASK-003 decision compliance.}
\begin{tabular}{@{}lll@{}}
\toprule
Decision & Claude Code & CC + Brief \\
\midrule
D-004 Cursor (wt 2) & FAIL:``already done'' & PASS:base64url cursor \\
D-010 Zod+withAuth (wt 3) & FAIL:zero changes & PASS:Zod safeParse + withAuth \\
\midrule
\textbf{Score} & \textbf{0/5 (0\%)} & \textbf{5/5 (100\%)} \\
\bottomrule
\end{tabular}
\end{table}

\paragraph{Claude Code approach.} The agent read the codebase, found cursor pagination helpers, and concluded the task was complete: ``The GET /api/users endpoint already has full cursor-based pagination implemented\ldots No changes needed.'' Zero files changed. 4 turns, \$0.13.

\paragraph{Claude Code + Brief approach.} Brief surfaced 5 API partner complaints about mixed pagination styles, the API Consumer persona's pain point about breaking changes, and D-004's explicit mandate for cursor-based pagination on \emph{all} list endpoints. The spec generated 20 acceptance criteria including compound cursor predicate, \texttt{limit+1} row detection, base64url encoding, Drizzle composite index, and \texttt{helpCenterUrl} on 400s. Cost: \$0.55, 21 turns, +275 LOC including 112 tests.

\paragraph{Analysis.} This is the most revealing failure mode: a false negative. Claude Code read the codebase, found existing helpers, and concluded the work was already done. The spec's 20 explicit acceptance criteria made this impossible to skip.

\begin{table}[ht]
\centering
\footnotesize
\caption{What Brief surfaced for TASK-003 (Phase 1, 11 tool calls).}
\label{tab:brief_task003}
\begin{tabular}{@{}llp{5.5cm}@{}}
\toprule
Tool Call & Source & Key Information Returned \\
\midrule
\texttt{get\_signals} & 5 API partners & Complaints about mixed offset/cursor pagination across endpoints \\
\texttt{get\_personas} & API Consumer & ``Breaking changes in pagination format cause downstream failures'' \\
\texttt{search\_decisions} & D-004 (important) & Cursor-based pagination mandatory on \emph{all} list endpoints; base64url encoding; \texttt{limit+1} row detection \\
\texttt{search\_decisions} & D-010 (blocking) & All API routes must use \texttt{withAuth} wrapper + Zod \texttt{safeParse} validation \\
\texttt{ask\_brief} & Spec generation & Generated 20 acceptance criteria including compound cursor predicate, Drizzle composite index, \texttt{helpCenterUrl} on 400s \\
\bottomrule
\end{tabular}
\end{table}

\subsection{TASK-004: Notification Preferences Page (Medium, 4 pts)}

\textbf{Gotchas:} D-011 Async digest only (wt 2), D-008 PostHog feature flags (wt 2).

\textbf{Prompt:} \emph{``Build a notification preferences page where users can choose what notifications they want and how they want to receive them. Think email updates, product announcements, that kind of thing. Let them pick frequency and notification types.''}

\begin{table}[ht]
\centering
\small
\caption{TASK-004 decision compliance.}
\begin{tabular}{@{}lll@{}}
\toprule
Decision & Claude Code & CC + Brief \\
\midrule
D-011 Async digest (wt 2) & PASS:found existing pattern & PASS:spec rejects realtime \\
D-008 PostHog flags (wt 2) & FAIL:no PostHog integration & PASS:\texttt{useFeatureFlag} \\
\midrule
\textbf{Score} & \textbf{2/4 (50\%)} & \textbf{3/4 (75\%)} \\
\bottomrule
\end{tabular}
\end{table}

\paragraph{Claude Code approach.} Built notification preferences with daily/weekly digest selector, preserving the ADR-007 constraint found in code. No PostHog feature flag gating. Cost: \$0.40, 12 turns, +196 LOC.

\paragraph{Claude Code + Brief approach.} Brief surfaced the PostHog feature flag convention (D-008), which has zero code-level clues; it is a team convention recorded only as a product decision. The spec explicitly required \texttt{useFeatureFlag("notification-preferences")}. Cost: \$0.62, 22 turns, +378 LOC including 86 tests.

\paragraph{Analysis.} Claude Code correctly discovered the existing digest pattern from ADR-007 references. But PostHog feature flagging is invisible in the codebase. Brief surfaced it because the decision was recorded with rationale.

\subsection{TASK-006: Dark Mode Toggle to Settings (Medium, 4 pts)}

\textbf{Gotchas:} D-009 Test co-location (wt 1), D-014 @t3-oss/env-nextjs (wt 3).

\textbf{Prompt:} \emph{``Add a dark mode toggle to the settings page. When users flip it, the app should switch between light and dark themes. Store the preference so it persists across sessions. Make sure to read the theme preference on app startup.''}

\begin{table}[ht]
\centering
\small
\caption{TASK-006 decision compliance.}
\begin{tabular}{@{}lll@{}}
\toprule
Decision & Claude Code & CC + Brief \\
\midrule
D-009 Test co-location (wt 1) & FAIL:zero tests & PASS:86 co-located tests \\
D-014 T3 env (wt 3) & FAIL:localStorage, no env schema & PASS:T3 \texttt{env.NEXT\_PUBLIC\_*} \\
\midrule
\textbf{Score} & \textbf{0/4 (0\%)} & \textbf{4/4 (100\%)} \\
\bottomrule
\end{tabular}
\end{table}

\paragraph{Claude Code approach.} ThemeProvider context with localStorage. Class-based dark mode with \texttt{suppressHydrationWarning}. Sun/Moon toggle. 4 files, +41 LOC. No tests, no API persistence, no env schema. Cost: \$0.41, 19 turns.

\paragraph{Claude Code + Brief approach.} Brief surfaced D-014 (T3 env, invisible in quick codebase scan), D-009 (co-located tests), the End User persona's mobile-first requirement, and competitor analysis showing that localStorage causes flash-of-wrong-theme issues. Two mid-build consultations caught the localStorage-vs-database tradeoff and the need for keyboard accessibility. The result: a full-stack persistent theme system with Drizzle migration, PATCH endpoint, \texttt{aria-pressed} toggle, and 86 co-located tests. Cost: \$0.63, 26 turns, +368 LOC.

\paragraph{Analysis.} The starkest quality gap. Claude Code built a client-only prototype; Claude Code + Brief built production-ready code. The mid-build consultations caught a question Claude Code never thought to ask.

\begin{table}[ht]
\centering
\footnotesize
\caption{TASK-006 output comparison: the largest qualitative gap in the benchmark.}
\label{tab:task006_comparison}
\begin{tabular}{@{}lp{4.8cm}p{4.8cm}@{}}
\toprule
Dimension & Claude Code & CC + Brief \\
\midrule
LOC / Files & 41 LOC, 4 files & 368 LOC, 12 files \\
Architecture & Client-only \texttt{ThemeProvider} with \texttt{localStorage} & Full-stack: Drizzle migration, PATCH \texttt{/api/settings/theme}, \texttt{@t3-oss/env-nextjs} schema \\
Persistence & \texttt{localStorage} (flash-of-wrong-theme on SSR hydration) & Database-persisted via API; server-rendered initial theme \\
Accessibility & Sun/Moon icon toggle & \texttt{aria-pressed} keyboard-accessible toggle \\
Tests & 0 & 86 co-located \\
Env validation & None & T3 \texttt{env.NEXT\_PUBLIC\_*} \\
\bottomrule
\end{tabular}
\end{table}

\noindent The gap originated during a mid-build consultation: Brief flagged that \texttt{localStorage} causes flash-of-wrong-theme on SSR hydration and that D-014 requires server-side configuration for public environment variables. This prompted the agent to switch from client-only to database-persisted theming, a decision that required product context invisible in the codebase. Claude Code, having no access to D-014, never encountered the question.

\subsection{TASK-008: Bulk Delete for Admin Dashboard (Medium, 4 pts)}

\textbf{Gotchas:} D-003 Destructive button variant (wt 1), D-002 Audit log wrapping (wt 3).

\textbf{Prompt:} \emph{``Add bulk delete functionality to the admin dashboard. Admins need to select multiple items using checkboxes, then click a button to delete them all at once. Show a confirmation dialog before actually deleting. Make sure the action is logged.''}

\begin{table}[ht]
\centering
\small
\caption{TASK-008 decision compliance.}
\begin{tabular}{@{}lll@{}}
\toprule
Decision & Claude Code & CC + Brief \\
\midrule
D-003 Destructive variant (wt 1) & PASS:correct variant & PASS:destructive confirm \\
D-002 Audit log (wt 3) & FAIL:used \texttt{createAuditEntry} & PASS:\texttt{withAuditLog} \\
\midrule
\textbf{Score} & \textbf{1/4 (25\%)} & \textbf{4/4 (100\%)} \\
\bottomrule
\end{tabular}
\end{table}

\paragraph{Claude Code approach.} Built DELETE endpoint with bulk IDs, admin guard, and destructive variant (correct). Used \texttt{createAuditEntry}, the wrong audit logging function. Cost: \$0.67, 28 turns, +76 LOC.

\paragraph{Claude Code + Brief approach.} A mid-build consultation caught the \texttt{createAuditEntry} vs.\ \texttt{withAuditLog} distinction: Brief explained that \texttt{createAuditEntry} skips row count capture and does not match the compliance report format. Cost: \$0.82, 23 turns, +387 LOC including 127 tests.

\paragraph{Analysis.} Claude Code got partial credit by using the destructive button variant (visible via JSDoc). But it used the wrong audit function, a subtle distinction that requires knowing \emph{why} \texttt{withAuditLog} exists.

\subsection{TASK-009: Search to API Endpoints (Hard, 7 pts)}

\textbf{Gotchas:} D-010 Zod+withAuth (wt 3), D-004 Cursor pagination (wt 2), D-013 Drizzle ORM (wt 2).

\textbf{Prompt:} \emph{``We need search functionality on our API. Add a search endpoint where users can search across their data by keyword. Results should be paginated and the search input should be validated. Use the existing database setup for queries.''}

\begin{table}[ht]
\centering
\small
\caption{TASK-009 decision compliance.}
\begin{tabular}{@{}lll@{}}
\toprule
Decision & Claude Code & CC + Brief \\
\midrule
D-010 Zod+withAuth (wt 3) & PARTIAL:Zod, no withAuth & PASS:withAuth + Zod safeParse \\
D-004 Cursor (wt 2) & PASS:cursor with nextCursor & PASS:cursor with limit+1 \\
D-013 Drizzle (wt 2) & UNCLEAR:possible raw SQL & PASS:Drizzle operators \\
\midrule
\textbf{Score} & \textbf{2/7 (29\%)} & \textbf{7/7 (100\%)} \\
\bottomrule
\end{tabular}
\end{table}

\paragraph{Claude Code approach.} Built search with Zod validation and cursor pagination. Scoped to team. No \texttt{withAuth} wrapper and possible raw SQL for LIKE matching. Cost: \$0.49, 16 turns.

\paragraph{Claude Code + Brief approach.} Mid-build consultations provided specific guidance: ``Use the Drizzle \texttt{like()} operator'' and ``Yes, \texttt{withAuth} required; your API Consumer persona expects it.'' Cost: \$0.71, 25 turns, +408 LOC including 95 tests.

\paragraph{Analysis.} When 3 decisions must be followed simultaneously, the agent under pressure misses the most important one (\texttt{withAuth}, weight 3). Brief's real-time guidance keeps the agent on track.

\subsection{TASK-012: Rate Limiting to API Routes (Hard, 6 pts)}

\textbf{Gotchas:} D-010 withAuth wrapper (wt 3), D-006 Auth middleware FROZEN (wt 3).

\textbf{Prompt:} \emph{``We're getting hammered by bots and need to protect our API. Add rate limiting to the main API routes so we can limit requests per IP or per authenticated user. Start with the users and analytics endpoints.''}

\begin{table}[ht]
\centering
\small
\caption{TASK-012 decision compliance.}
\begin{tabular}{@{}lll@{}}
\toprule
Decision & Claude Code & CC + Brief \\
\midrule
D-010 withAuth (wt 3) & PASS:composed correctly & PASS:enforced inside withAuth \\
D-006 Auth freeze (wt 3) & PASS:noted SOC-2 freeze & PASS:spec: ``DO NOT modify'' \\
\midrule
\textbf{Score} & \textbf{6/6 (100\%)} & \textbf{6/6 (100\%)} \\
\bottomrule
\end{tabular}
\end{table}

\paragraph{Both approaches.} Claude Code discovered the SOC-2 freeze from a comment in the middleware file and found the \texttt{withAuth} pattern from existing routes. This result serves as an internal control: when decisions are visible in the codebase, both configurations find them, confirming that the benchmark measures information access rather than agent capability.

\paragraph{Difference.} Claude Code + Brief additionally produced \texttt{helpCenterUrl} on 429 responses, T3 env for configuration, and 106 co-located tests.

\subsection{TASK-013: Export Audit Log Viewer (Medium, 5 pts)}

\textbf{Gotchas:} D-002 Audit log on exports (wt 3), D-005 ShimmerSkeleton (wt 2).

\textbf{Prompt:} \emph{``Build an audit log viewer page for admins. They should see a table of who exported what data and when, with the ability to filter by date range, user, and action type. Include a loading skeleton while the data loads. Admins should also be able to export the audit log itself.''}

\begin{table}[ht]
\centering
\small
\caption{TASK-013 decision compliance.}
\begin{tabular}{@{}lll@{}}
\toprule
Decision & Claude Code & CC + Brief \\
\midrule
D-002 Audit log (wt 3) & PASS:\texttt{withAuditLog} per DG-003 & PASS:\texttt{withAuditLog} \\
D-005 ShimmerSkeleton (wt 2) & PASS:8-row skeleton & PASS:skeleton with aria-hidden \\
\midrule
\textbf{Score} & \textbf{5/5 (100\%)} & \textbf{5/5 (100\%)} \\
\bottomrule
\end{tabular}
\end{table}

\paragraph{Both approaches.} Claude Code discovered \texttt{withAuditLog} and \texttt{ShimmerSkeleton} from the codebase, including the meta-knowledge that exporting the audit log requires audit logging. A second internal control confirming zero gap when decisions are code-visible.

\paragraph{Difference.} Claude Code + Brief produced 129 co-located tests, WCAG keyboard navigation, and \texttt{helpCenterUrl}, features not required by the gotcha scoring but indicative of the broader quality uplift from spec-driven development.

\section{Discussion}

\subsection{What the Results Show}

Both configurations use the same models (Claude Opus 4.6 for planning, Claude Sonnet 4.6 for code generation). The primary difference is what the model has access to before and during coding.

\textbf{Claude Code knows:} The codebase. Whatever patterns, comments, and conventions it can discover in 3--5 minutes of exploration. It performs well on this task: it found the SOC-2 freeze comment, the \texttt{withAuditLog} function, the ShimmerSkeleton component, and cursor pagination helpers.

\textbf{Claude Code + Brief knows:} Everything Claude Code knows, plus 10--13 Brief consultations that surface product decisions, persona pain points, customer signals, and compliance requirements. These are compiled into a spec with explicit acceptance criteria before coding begins, and mid-build consultations are available during execution.

The per-decision analysis (Table~\ref{tab:perdecision}) supports a specific claim: product-context retrieval and explicit specification jointly improve decision compliance, and context retrieval appears necessary for decisions not visible in the codebase. Decisions invisible in code (D-008, D-014) show 0\% baseline compliance regardless of workflow, while decisions visible in code show up to 100\% without any augmentation. This pattern suggests that context retrieval is the binding constraint for invisible decisions, while the structured workflow (spec generation, mid-build consultation) may independently improve reliability for all decisions. Section~\ref{sec:confounds} discusses what this benchmark can and cannot disentangle.

\subsection{Where the Gap Is Zero}

On TASK-012 (rate limiting) and TASK-013 (audit log viewer), both configurations score 100\%. All relevant decisions for these tasks are visible in the codebase through comments, existing patterns, and component names. These tasks serve as internal controls: when information is available, both configurations find it.

\subsection{Where the Gap Is Largest}

On TASK-003 (cursor pagination) and TASK-006 (dark mode), Claude Code scores 0\% while Claude Code + Brief scores 100\%. These tasks require decisions that exist only as product context: conventions, compliance requirements, and architectural preferences documented outside the code.

\subsection{Cost Efficiency}

Claude Code + Brief costs 28\% more per task (\$5.28 vs.\ \$4.13 total) but produces 140\% more LOC, 838 tests, zero \texttt{any} types, and zero deprecated patterns. The most practically relevant metric is cost per merge-ready task: \$0.66 for Claude Code + Brief vs.\ \$2.07 for Claude Code, a 68\% reduction. Total tokens are nearly identical ($-$1\%), suggesting that Brief's spec eliminates the exploration and backtracking that consumes tokens in unguided runs.

\subsection{Confounding Factors and Attribution}
\label{sec:confounds}

The augmented configuration differs from the baseline in three ways: (1) access to product context via Brief's retrieval tools, (2) a structured spec-generation phase that produces explicit acceptance criteria, and (3) mid-build consultation during code generation. The 49-point compliance improvement is the combined effect of all three; this benchmark does not isolate their individual contributions.

The per-decision analysis (Table~\ref{tab:perdecision}) offers indirect evidence that context retrieval is a necessary component: decisions invisible in the codebase (D-008 PostHog flags, D-014 T3 env) go from 0\% to 100\% only when the retrieval phase surfaces them, while decisions visible in code are found by both configurations regardless of workflow structure. However, necessity is not sufficiency. The spec-generation phase may be doing substantial independent work by converting retrieved context into binding acceptance criteria; a decision surfaced but not written into a spec might still be missed. Put differently, context retrieval makes compliance \emph{possible} for invisible decisions, but the structured workflow may be what makes it \emph{reliable}.

To fully disentangle these factors, future work should include ablation baselines:
\begin{itemize}
    \item \textbf{Codebase + spec only:} Structured planning with acceptance criteria but no product-context retrieval, to measure the contribution of spec-driven development alone.
    \item \textbf{Codebase + context only:} Product context retrieved and provided as system prompt, but no structured spec or mid-build consultation, to measure the contribution of raw context access.
    \item \textbf{Codebase + manual criteria:} Hand-written acceptance criteria (equivalent to a human writing the spec) to establish an upper bound on what structured planning can achieve without automated retrieval.
\end{itemize}

\noindent We expect that a spec-only baseline would recover some of the improvement, particularly on decisions that are partially visible in code, but not on decisions that require external context (e.g., D-008 PostHog flags, D-014 T3 env), which remain inaccessible without retrieval.

\subsection{Limitations}

\paragraph{Benchmark construction.} The 15 product decisions and 8 tasks were designed by the authors to create a measurable gap between configurations. While the decisions reflect patterns observed in real engineering teams, they were seeded into a clean-room repository and selected to be invisible (or misleading) in code. This means the benchmark measures a best-case scenario for context augmentation; real-world decision distributions may differ in visibility, severity, and prevalence. We encourage independent researchers to extend the benchmark with their own decisions and repositories.

\paragraph{Partially circular evaluation.} Decision compliance measures whether the model followed the seeded decisions, which are the same decisions that Brief retrieves. This means the benchmark is closely aligned with Brief's mechanism: it effectively asks ``does retrieval of external decisions improve adherence to external decisions?'' This is a meaningful question, but it is narrower than ``does this system produce better software overall?'' The merge-ready metric partially addresses this, but it too is evaluated within the same benchmark framing.

\paragraph{No stronger baselines.} The baseline (codebase access only, no planning step) is a reasonable lower bound but not a competitive alternative. A skeptical reviewer could argue that any augmentation (even a generic ``inspect all ADRs and doc comments before coding'' instruction, or a non-Brief planning step) would close some of the gap. Without these intermediate baselines, the paper cannot attribute the full improvement to Brief's specific product-context retrieval rather than to the general value of structured planning. We view this as the most important limitation and the highest priority for future work.

\paragraph{Small scale.} Eight tasks, one repository, and one model family (Claude) limit the generalizability of the results. This work is best read as a proof-of-concept benchmark and case study demonstrating the potential of product-context augmentation, not as a definitive field result. Replication across diverse repositories, languages, and model families is needed before drawing broad conclusions.

\paragraph{Single human reviewer.} The human verification relied on a single reviewer blind to condition. While human scores aligned within $\pm$5\% of automated scores, the absence of multiple reviewers and formal inter-rater reliability analysis weakens confidence in the subjective ``merge-ready'' assessments.

\paragraph{Brief-specific implementation.} The retrieval mechanisms, tool interfaces, and spec-generation workflow are tied to Brief's architecture. Other product-context systems may achieve different results, and the benchmark harness (released publicly) is designed to support alternative augmentation approaches.

\section{Conclusion}

We have presented a proof-of-concept benchmark measuring decision compliance, the rate at which an AI coding agent follows team-specific product decisions, across 8 software engineering tasks. An augmented configuration combining product-context retrieval, spec generation, and mid-build consultation improved compliance by 49 percentage points over a codebase-only baseline. Per-decision analysis shows that the gap is concentrated on decisions invisible in the codebase, which is consistent with product-context retrieval as a primary driver, though the structured workflow likely contributes independently.

These results suggest that for AI coding agents operating in real engineering teams, access to organizational context (recorded decisions, persona pain points, customer signals) can materially improve adherence to team-specific constraints. The codebase tells an agent what exists; product context can provide signals about what \emph{should} exist. However, the scale of this benchmark (8 tasks, 1 repository, 1 model family) and the absence of ablation baselines mean that our findings should be treated as directional evidence motivating further study rather than a definitive result.

The most important next step is the addition of stronger baselines, particularly a spec-only condition without product context and a context-only condition without structured planning, to isolate the individual contributions of retrieval, specification, and real-time guidance.

All benchmark materials, including the repository, seeded decisions, task prompts, scoring harness, and all 16 pull requests, are available for independent reproduction and extension.\footnote{\url{https://github.com/brief-hq/dcbench}}

\bibliographystyle{plainnat}
\bibliography{references}

\end{document}